\documentclass{PoS}

\title{Hadronic corrections to the muon anomalous magnetic moment from lattice QCD}

\ShortTitle{Hadronic corrections to the muon $g-2$ from lattice QCD }

\author{\speaker{T. Blum}
       \\
        University of Connecticut and the RIKEN BNL Research Center\\
        E-mail: \email{tblum@phys.uconn.edu}}
\author{M. Hayakawa\\
        University of Nagoya\\
       E-mail: \email{hayakawa@eken.phys.nagoya-u.ac.jp}}
\author{T. Izubuchi\\
        Brookhaven National Laboratory and RIKEN BNL Research Center\\
       E-mail: \email{izubuchi@bnl.gov}}

\abstract{After a brief self-contained introduction to the muon anomalous magnetic moment, $(g-2)_\mu$, we review the status of lattice calculations of the hadronic vacuum polarization contribution and present first results from lattice QCD for the hadronic light-by-light scattering contribution. The signal for the latter is consistent with model calculations.  While encouraging, the statistical error is large and systematic errors are mostly uncontrolled. The method is applied first to pure QED as a check. }

\FullConference{The 30 International Symposium on Lattice Field Theory - Lattice 2012,\\
		June 24-29, 2012\\
		Cairns, Australia}

\begin{document}

\section{Introduction}
\begin{figure}[htbp]
\begin{center}
\includegraphics[width=0.7\textwidth]{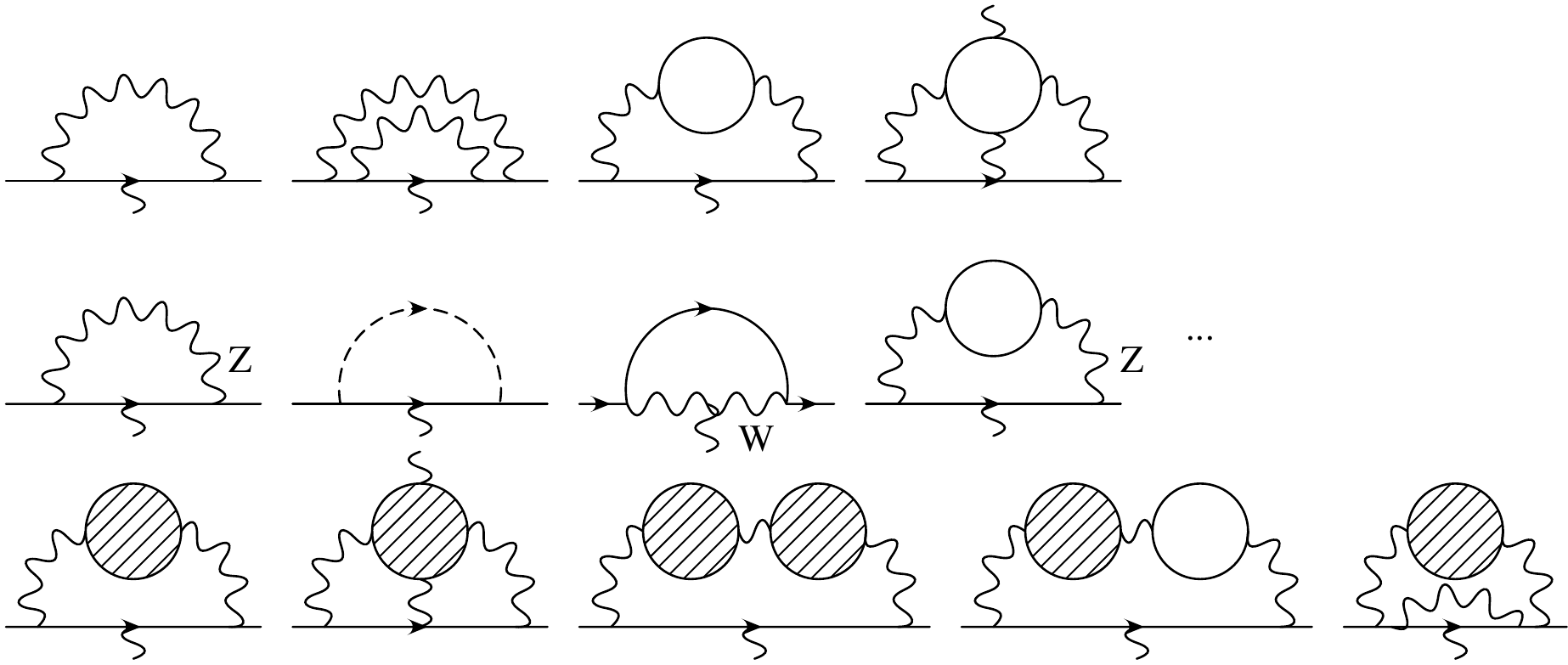} 
\caption{Representative diagrams, up to order $\alpha^{3}$, in the Standard Model that contribute to the muon anomaly. The rows, from to top to bottom, correspond to QED, EW, and QCD. Horizontal solid lines represent the muon, wiggly lines denote photons unless otherwise labeled, other solid lines are leptons, filled loops denote quarks (hadrons), and the dashed line represents the higgs boson.}
\label{fig:rad}
\end{center}
\end{figure}

The free Dirac equation predicts that an elementary fermion has an intrinsic magnetic moment proportional to its spin $\vec{S}$,
 \begin{eqnarray}
\vec{\mu} &=& g\,\left(\frac{e}{2m}\right)\, \vec{S},
\end{eqnarray}
where the Land\'e $g$-factor is exactly 2, $e$ is the fundamental electric charge, and $m$ the fermion's mass ($\hbar=c=1$). In the presence of interactions, $g$ receives radiative corrections (Fig.~\ref{fig:rad}), and these can be computed order-by-order in weak coupling perturbation theory in the electromagnetic coupling $\alpha\equiv e^{2}/4 \pi$. As discussed below, in addition, the hadronic corrections require non-perturbative methods or experimental data.

The fundamental vector interaction of the fermion with an external electromagnetic field is given by the matrix element of the electromagnetic current between incoming and outgoing fermion states of momentum and spin $\{p,s\}$ and $\{p',s'\}$,
\begin{eqnarray}
\left \langle p',s'|\bar\psi \gamma_\mu\psi|p,s\right\rangle &=&
 \bar u(p',s') \left (F_1(Q^2)\gamma_\mu +  i \frac{F_2(Q^2)}{4 m}[\gamma_\mu,\gamma_\nu] Q_\nu \right)u(p,s),
\label{eq:ff}
\end{eqnarray}
where $u(p,s)$, $\bar u(p',s')$ are spinors, and $Q_\nu=(p-p')_\nu$ is the (space-like) momentum transferred to the fermion by the photon. The form of Eq.~(\ref{eq:ff}) is dictated by Lorentz-invariance, $P$ and $C$ symmetries, and the Ward-Takahashi identity; hence the form factors $F_1(Q^2)$ and $F_2(Q^2)$ contain all information on the fermion's interaction with the photon. In particular, in the limit $Q^2\to0$, $F_1(0)$ is the charge of the fermion in units of $e$, and
\begin{eqnarray}
F_2(0) &=& \frac{g-2}{2}
\end{eqnarray}
is the anomalous magnetic moment. In the following $F_2(0)\equiv a_\mu$ is referred to as the muon "anomaly". 

The muon anomaly provides one of the most stringent tests of the Standard Model because it has been measured to fantastic accuracy (0.54ppm)~\cite{Bennett:2006fi} and calculated to even better precision~\cite{Aoyama:2012wk,Davier:2010nc,Hagiwara:2011af}. The difference between the two is reported to range between $249(87)\times 10^{-11}$ and $287(80)\times 10^{-11}$, or about 2.9 to 3.6 standard deviations~\cite{Aoyama:2012wk,Davier:2010nc,Hagiwara:2011af}. The Standard Model precision is attained by calculating contributions (in orders of $\alpha$) from QED ($\alpha^5$)~\cite{Aoyama:2012wk}, electroweak (EW) ($\alpha^2$)~\cite{Czarnecki:2002nt}, and QCD ($\alpha^3$). The latter represent the largest uncertainty. The lowest order ($\alpha^2$) contribution arises from the hadronic vacuum polarization (HVP), $\Pi(Q^2)$, while the next-to-leading contribution is from hadronic light-by-light (HLbL) scattering and the HVP accompanied by  an additional photon.  The various contributions to $a_\mu$ are summarized in Tab.~\ref{tab:summary} and depicted in Fig.~\ref{fig:rad}. It is well known that the discrepancy between theory and experiment is large in magnitude, roughly two times the EW contribution, and has important consequences for new physics. Because of its central role in constraining beyond the Standard Model (BSM) theories, the muon anomaly will be measured to even better accuracy in new experiments at Fermilab (E989) and J-PARC (E34). The goal of E989 is to reduce the total error to 0.14 ppm. A brief summary of the importance of $(g-2)_\mu$ to BSM physics can be found in~\cite{Hewett:2012ns}.

\begin{table}[htdp]
\caption{Standard Model contributions to the muon anomaly. The QED contribution is through $\alpha^5$, EW $\alpha^2$, and QCD $\alpha^3$. The two QED values correspond to different values of $\alpha$, and QCD to lowest order (LO) contributions from the hadronic vacuum polarization (HVP) using $e^+e^-\to\rm hadrons$ and $\tau\to\rm hadrons$, higher order (HO) from HVP and an additional photon, and hadronic light-by-light (HLbL) scattering.}
\begin{center}
\begin{tabular}{c|lll}
\hline
QED & &$116\,584\,71.8\,845\,(9)(19)(7)(30) \times 10^{-10}$ & \cite{Aoyama:2012wk}\\
        & &$116\,584\,71.8\,951\,(9)(19)(7)(77) \times 10^{-10}$ & \cite{Aoyama:2012wk}\\
        \hline
EW &  & $15.4\,(2)\times 10^{-10}$ & \cite{Czarnecki:2002nt}\\
\hline
QCD & LO ($e^+e^-$) & $692.3\,(4.2)\times 10^{-10}$, 
	   $694.91\,(3.72)\,(2.10)\times10^{-10}$ & \cite{Davier:2010nc,Hagiwara:2011af}\\
          & LO ($\tau$) & $701.5\,(4.7)\times 10^{-10}$ & \cite{Davier:2010nc}\\
          & HO HVP & $-9.79 (9)\times 10^{-10}$ &\cite{Hagiwara:2006jt}\\
          & HLbL & $10.5(2.6)\times 10^{-10}$ & \cite{Prades:2009tw}\\
\hline
\end{tabular}
\end{center}
\label{tab:summary}
\end{table}%

The HVP contribution to the muon anomaly has been computed using the experimentally measured cross-section for the reaction $e^+e^-\to$ hadrons and a dispersion relation to relate the real and imaginary parts of $\Pi(Q^2)$. The current quoted precision on such calculations is a bit more than one-half of one percent~\cite{Davier:2010nc,Hagiwara:2011af}. The HVP contributions can also be calculated from first principles in lattice QCD~\cite{Blum:2002ii}. While the current precision is significantly higher for the dispersive method, lattice calculations are poised to reduce errors significantly in next one or two years. These will provide important checks of the dispersive method before the new Fermilab experiment.
Unlike the case for $a_\mu(\rm HVP)$, $a_\mu(\rm HLbL)$ can not be computed from experimental data and a dispersion relation (there are many off-shell form factors that enter which can not be measured). While model calculations exist (see~\cite{Prades:2009tw} for a summary), they are not systematically improvable. A determination using lattice QCD where all errors are controlled is therefore desirable.

In Sec.~\ref{sec:hvp} we review the status of lattice calculations of $a_\mu(\rm HVP)$. Section~\ref{sec:hlbl} is a presentation of our results for $a_\mu(\rm HLbL)$ computed in the framework of lattice QCD+QED. Section~\ref{sec:con} gives our conclusions and outlook for future calculations.

\section{Hadronic Vacuum Polarization}
\label{sec:hvp}

It has long been known that $a_\mu(\rm HVP)$ can be computed from $e^+e^-\to $ hadrons~\cite{Durand:1962zzb,Gourdin:1969dm}. Later it was realized that $\tau\to$ hadrons could also be used to improve the result in a limited, but important, range of $Q^2$~\cite{Alemany:1997tn}. For example, the most recent studies find \cite{Hagiwara:2011af} $a_\mu(\rm HVP) =(694.91\pm3.72\pm2.10)\times10^{-10}$ ($e^+e^-$) while~\cite{Davier:2010nc} quotes $a_\mu(\rm HVP)=(692\pm4.2)\times 10^{-10}$ ($e^+e^-$) and $(701.5\pm4.7)\times 10^{-10}$ (including $\tau$). These determinations are the targets for lattice calculations. Note the result including $\tau$ data is about 2 standard deviations larger than the pure $e^+e^-$ contribution, and reduces the discrepancy with the Standard Model to 2.4 standard deviations~\cite{Davier:2010nc}. The former requires isospin corrections which may not be under control. Alternatively, $\rho-\gamma$ mixing may explain the difference and bring the $\tau$-based result in line with that of $e^+e^-$~\cite{Jegerlehner:2011ti}. The lattice QCD calculations, as we discuss below, are not yet at the level of precision to help resolve this issue.

\begin{figure}[htbp]
\begin{minipage}{0.45\textwidth}
\includegraphics[width=0.9\textwidth]{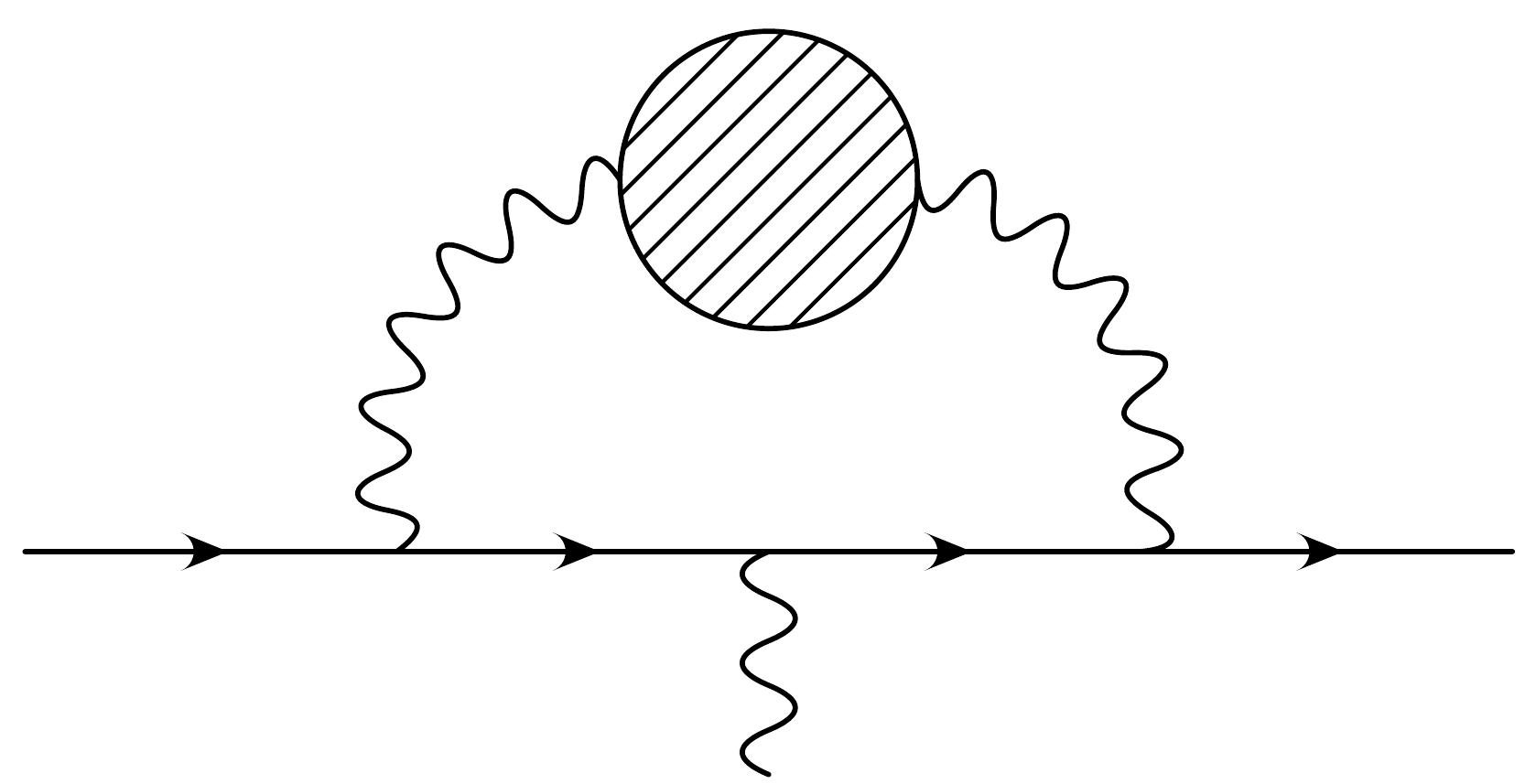} 
\caption{The lowest order hadronic contribution to the muon anomaly. The blob represents all possible intermediate hadronic states.}
\label{fig:lohvp}
\end{minipage}\hskip .1\textwidth
\begin{minipage}{0.45\textwidth}
\includegraphics[width=0.9\textwidth]{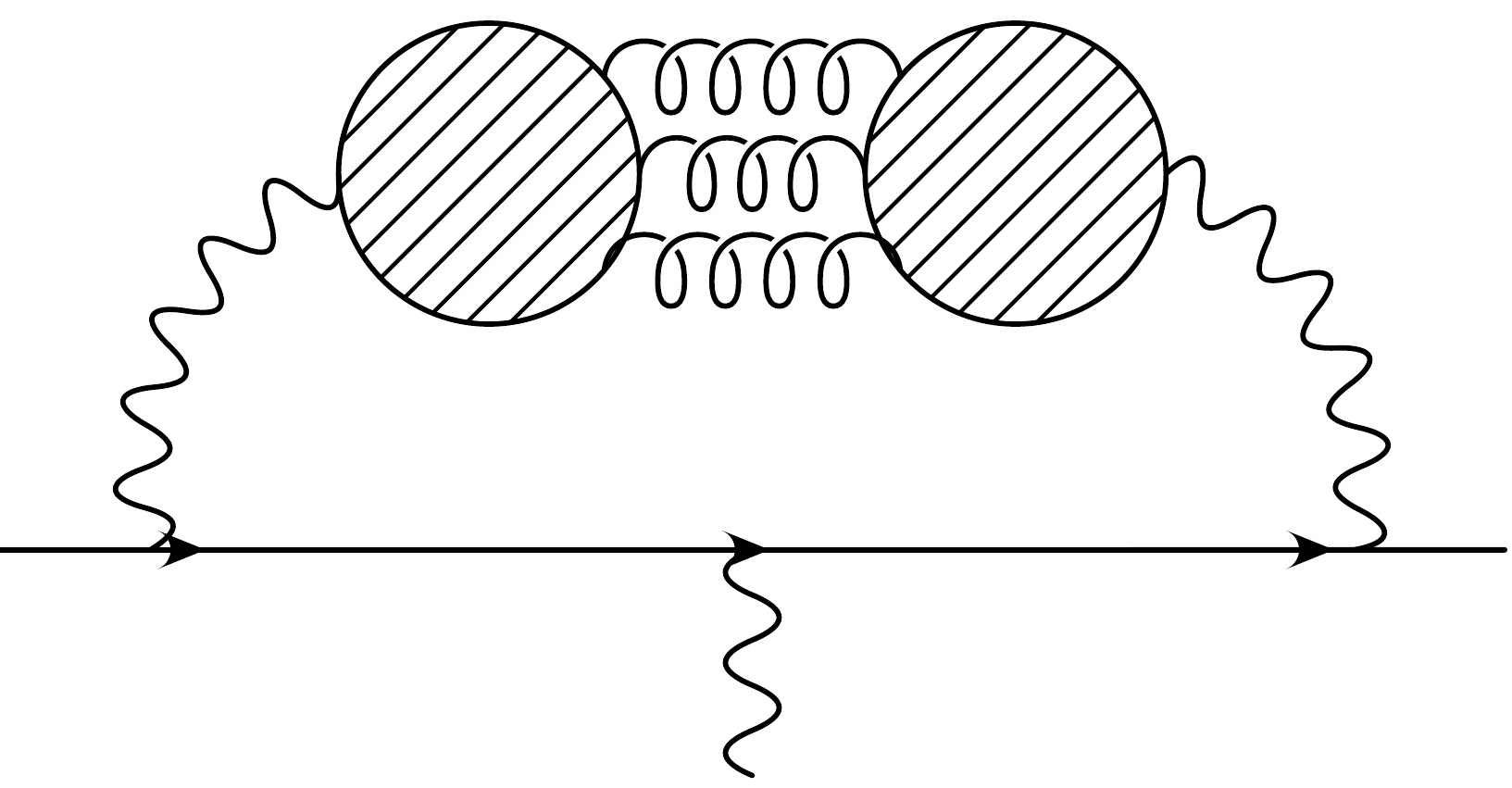} 
\caption{ Lowest order hadronic contribution to the muon anomaly from two quark loops connected by gluons. The contribution is flavor-symmetry and Zweig suppressed. }
\label{fig:lohvp disc}
\end{minipage}
\end{figure}

In~\cite{Blum:2002ii} it was shown that $a_\mu(\rm HVP)$ can be calculated instead using lattice QCD,
\begin{eqnarray}
a_\mu(\rm HVP) &=&\left(\frac{\alpha}{\pi}\right)^2 \int_0^\infty dQ^2 f(Q^2) \hat\Pi(Q^2).
\label{eq:latamu}
\end{eqnarray}
(a similar formula was found much earlier in the context of large N QCD~\cite{Lautrup:1971yp}).
Here $Q^2$ is the Euclidean momentum-squared in the one-loop QED integral (Fig.~\ref{fig:lohvp}) that runs through the renormalized (once-subtracted) vacuum polarization, $\hat\Pi(Q^2)$, and $f(Q^2)$ is a known analytic function that diverges in the limit $Q^2\to0$. Hence the integrand is dominated by the low momentum region around $m_\mu^2$.  The HVP is non-perturbative in this region, so the lattice provides a natural framework for the calculation of $a_\mu(\rm HVP)$. The importance of the low momentum region (and light quark masses) implies large volumes are necessary. Since lattice results are determined at discrete values of $Q^2$, a smooth interpolation is necessary in order to utilize Eq.~(\ref{eq:latamu}). $\hat\Pi(Q^2)$ is also cut-off at $|Q|\sim1/a$, so the high-momentum region can be handled with continuum perturbation theory~\cite{Chetyrkin:1996cf}, though this contribution is small~\cite{Blum:2002ii}, on the order of a percent for present lattice calculations.

First we briefly review the published lattice results for $a_\mu(\rm HVP)$ which are summarized in Tab.~\ref{tab:lathvp} (and have not changed since last year's meeting~\cite{Renner:2012fa}). We then update them from this year's meeting and discuss new developments. We discuss only results using dynamical fermions as it is known that the quenched results significantly underestimate the dispersive ones~\cite{Blum:2002ii,Gockeler:2003cw}. The quoted errors on all of the results are still somewhat crude, and nowhere complete.  The first dynamical results were obtained by Aubin and Blum using 2+1 flavors of improved staggered quarks~\cite{Aubin:2006xv}. The statistical errors are 2-3\%, but systematic errors are not controlled: there is only one lattice spacing so far, and the fits to $\Pi(Q^2)$, which are based on a resonance model+chiral perturbation theory for staggered quarks, are very sensitive to the low-momentum region. In addition, the vector mass in the pole is fixed to a value determined in a separate analysis. While a strength of the calculation is that Goldstone pion masses as light as $m_\pi\approx220$ MeV are used, the simple linear and quadratic extrapolations in the quark mass are problematic since the $\rho$-meson is stable in the calculation (as is true in all current lattice calculations, though at least one calculation where the $\rho$ is unstable is underway~\cite{Aubin:2012}).

The UKQCD group has computed $a_\mu(\rm HVP)$ using 2+1 flavors of domain wall fermions with several lattice spacings, volumes, and quark masses (down to $m_\pi\approx170$ MeV)~\cite{Boyle:2011hu}. They also developed a momentum fit based on two vector-particle poles in $\Pi(Q^2)$. In order to stabilize the fit, the lowest pole is fixed to the vector mass determined on the same lattices but from a separate analysis. The lattice spacing dependence can not be resolved outside relatively large statistical errors, so all results are fit together in a lattice-spacing-independent fit. An alternative quark mass fit~\cite{Feng:2011zk} provides an estimate of the chiral extrapolation error which is about 5\%, the same size as the statistical error.

The Mainz group uses two flavors of Wilson fermions with several volumes, relatively heavy pion masses (down to 280 MeV), a quenched strange quark, and several lattice spacings, but quotes results for $a_\mu(\rm HVP)$ from only one lattice spacing~\cite{DellaMorte:2011aa}. A partial systematic error of about 10\% coming from cut-off effects is estimated by re-analyzing the data with the lattice spacing shifted by one statistical standard deviation. The alternative chiral exptrapolation in ~\cite{Feng:2011zk} is also tried here, and no significant difference is found. The statistical error is also about 10\%. An innovation first used here that will likely be used in future calculations is twisted-boundary conditions for the quarks, so lower values of the momenta than allowed by periodic boundary conditions are accessible. 

Finally, the ETM collaboration has computed $a_\mu(\rm HVP)$ in two flavor QCD with twisted-mass Wilson fermions~\cite{Feng:2011zk}. They employ several lattice spacings, relatively small volumes, and heavy pion masses. The alternative chiral extrapolation mentioned above was proposed here and used to quote central values: The kernel in Eq.~(\ref{eq:latamu}) is rescaled by a reference hadronic observable, {\it e.g.}, $m_V$ or $f_V$. Nominally, this is designed to lead to smaller quark mass dependence, but in this case simply yields an overall shift in the data towards larger values of $a_\mu(\rm HVP)$ since the quark masses used are far from the chiral regime. As in the other studies, effects of volume and lattice spacing are not detectable within relatively large statistical errors, so all data is included in a lattice-spacing and volume independent fit. The final quoted error is simply the statistical uncertainty from this fit and is about 3\%.

\begin{table}[htdp]
\caption{Published lattice QCD results for $a_\mu(\rm HVP)(\times 10^{10}$).}
\begin{center}
\begin{tabular}{lllll}
\hline
$a_\mu$ & $N_f $  & errors & action  & group \\
\hline
713(15) & 2+1 & stat. & Asqtad & \cite{Aubin:2006xv}\\
748(21) & 2+1 & stat. & Asqtad & \cite{Aubin:2006xv}\\
641(33)(32)& 2+1 & stat., sys. & DWF & \cite{Boyle:2011hu} \\
572(16) & 2 & stat. & TM & \cite{Feng:2011zk} \\
618(64) & 2+1\footnote{strange quark is quenched} & stat., sys. & Wilson & \cite{DellaMorte:2011aa}   \\
\hline
\end{tabular}
\end{center}
\label{tab:lathvp}
\end{table}%

At this year's conference, two of the above groups presented preliminary results of improved calculations and methods. Jager presented results for the Mainz group with updated statistics and lighter quark masses ($m_\pi \ge 195$ MeV)~\cite{DellaMorte:2012cf}. Hotzel presented preliminary 2+1+1 results from the ETM collaboration~\cite{Feng:2012gh} in which the lightest pion mass is down to about 270 MeV. From the dispersive calculations the contribution from the charm quark is estimated to be about the same as the HLbL contribution, $a_\mu({\rm HVP,~charm})\approx a_\mu(\rm HLbL)\approx 10\times 10^{-10}$, and they find about double that amount.

\begin{figure}[htbp]
\begin{minipage}{0.45\textwidth}
\includegraphics[width=1.0\textwidth]{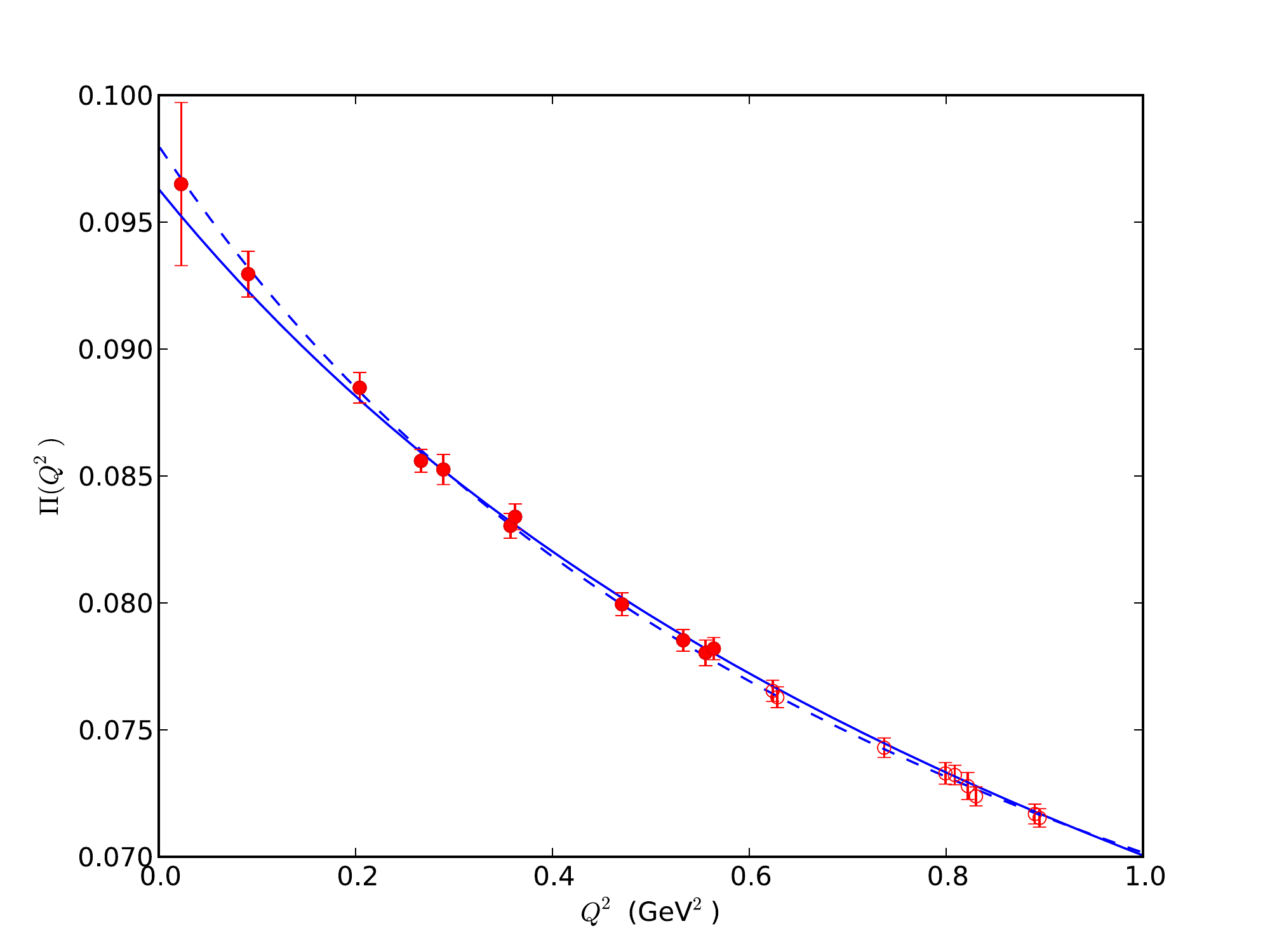} 
\caption{The hadronic vacuum polarization computed on a 2+1 flavor Asqtad ensemble from the MILC collaboration~\cite{Aubin:2012me,Golterman}. The solid line denotes a fit to the [1,1] Pad\'e approximant while the dashed line is from a vector-meson-dominance fit. }
\label{fig:qsq fit}
\end{minipage}\hskip .1\textwidth
\begin{minipage}{0.45\textwidth}
\includegraphics[width=0.9\textwidth]{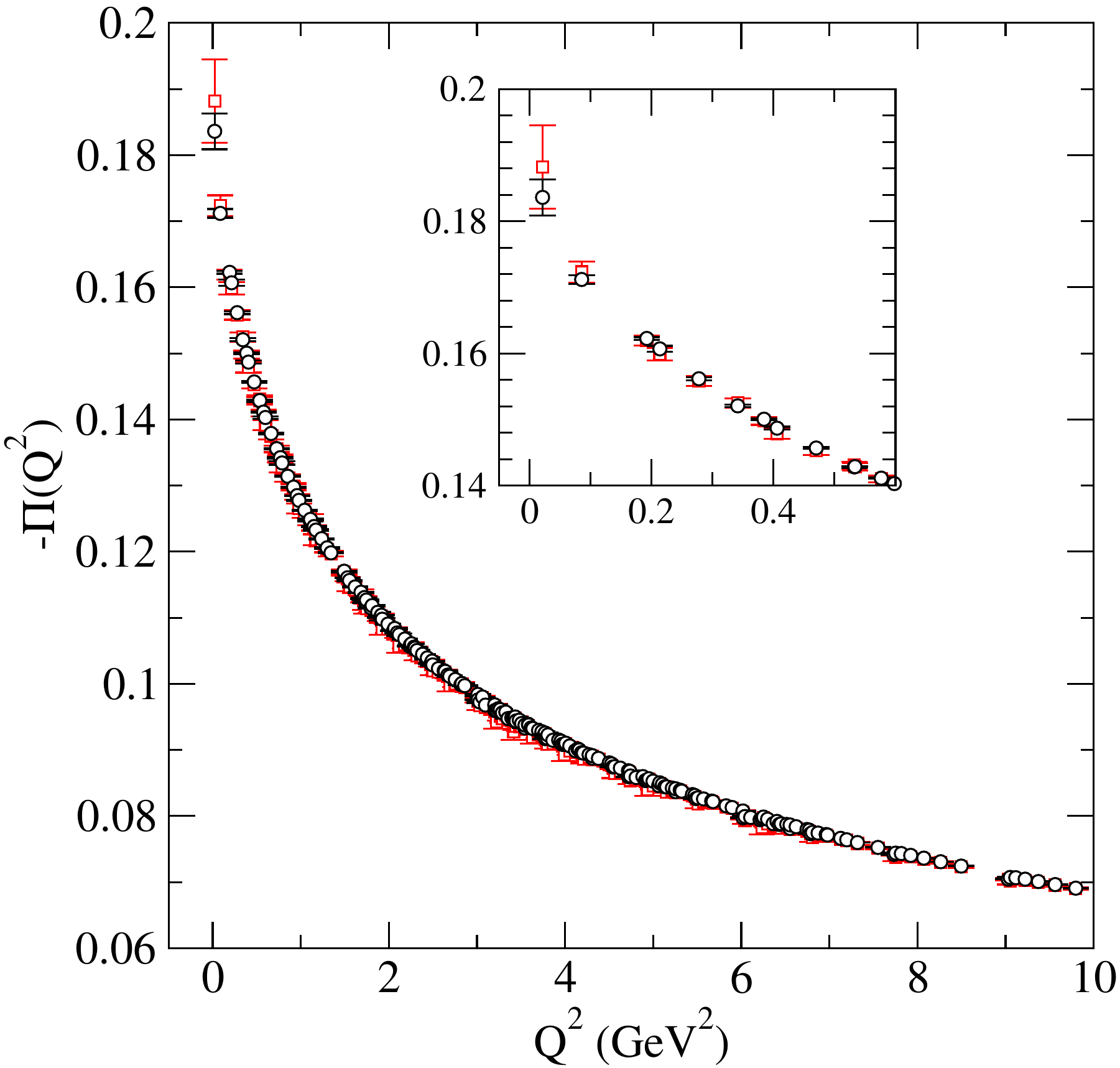} 
\caption{The hadronic vacuum polarization for a single flavor computed on a 2+1 flavor Asqtad ensemble from the MILC collaboration ($m_\pi=315$ MeV) using AMA (circles)~\cite{Blum:2012uh}. Original HVP calculation (squares) from~\cite{Aubin}. }
\label{fig:hvpAMA}
\end{minipage}
\end{figure}

Golterman presented a new model-independent method for fitting the $Q^2$ dependence of $\Pi(Q^2)$ based on Pad\'e approximants~\cite{Aubin:2012me,Golterman}. The central idea is that since $\hat\Pi(Q^2)$ is given by a positive spectral function, it is a so-called Steiljes function, obeying certain constraints, and can be obtained from a converging series of Pad\'e approximates evaluated on a sequence of $Q^2$ points. The Pad\'e approximates furnish a natural, model-independent, fit function for $\Pi(Q^2)$. Unfortunately, it is found that the new method disagrees outside of statistical errors with the vector-meson-dominance (VMD) fit function used earlier in~\cite{Aubin:2006xv}. This discrepancy can be traced to the aforementioned sensitivity of $a_\mu(\rm HVP)$ to the low momentum region. In Fig.~\ref{fig:qsq fit} we show the two fits which lead to $a_\mu(\rm HVP)=350(8)$ and 413(8), respectively. Both fits are good in the sense that they go through the data points and have acceptable $\chi^2$ values (the VMD fit is uncorrelated, {\it i.e.} the covariance matrix is diagonal), yet they give quite different results. The statistical errors on $\Pi(Q^2)$ are not small enough to differentiate the two fits, leading to an uncomfortably large systematic error. To make progress they must be reduced. In general, using higher $Q^2$ points where the errors are smaller leads to systematically low fits in the low momentum region~\cite{Aubin:2006xv}. Relatedly, uncorrelated fits or fits with more free parameters
mitigate the effect, but at the cost of larger statistical errors.
 
One way to reduce the statistical errors on $\Pi(Q^2)$ in the critical low momentum region is to use twisted boundary conditions to fill in this region~\cite{DellaMorte:2011aa,DellaMorte:2012cf}. This is not enough however, because the errors on each point are still too large. A new method described at the meeting by Shintani~\cite{Shintani,Blum:2012uh} that addresses this problem goes under the name "All Mode Averaging" (AMA). In Fig.~\ref{fig:hvpAMA} we show the HVP computed using the AMA technique. Computer resources required to obtain the same statistical error on $\Pi(Q^2)$ in the range $0\le Q^2\le 1.0$ GeV$^2$ were reduced by factors of 2.6-20 over the conventional method. AMA should be even more effective as the quark mass is reduced and the volume is increased.

So far we have only discussed the contribution to $a_{\mu}(\rm HVP)$ arising from the diagram shown in Fig.~\ref{fig:lohvp}, {\it i.e.,} from a single quark loop. In 2+1 flavor QCD and the $SU(3)$ flavor-symmetric limit, this is all there is. However flavor or isospin symmetry breaking gives rise to contributions from diagrams with two quark loops connected by gluons, as shown in Fig.~\ref{fig:lohvp disc}. While this contribution is flavor-symmetry and Zweig suppressed, it is expected to contribute on the 1\% level, possibly more, so must eventually be included. In~\cite{DellaMorte:2010aq}, using chiral perturbation theory, the two-quark-loop (``disconnected") diagram was shown to account for 10\% of the connected contribution. However, it is known that the pion contribution is only a small part of the total~\cite{Aubin:2006xv}. The two-quark-loop contribution was calculated in~\cite{Feng:2011zk}, but with a statistical precision that was not significant.

\section{Light by Light Amplitude}
\label{sec:hlbl}

\begin{figure}[htbp]
\begin{minipage}{0.45\textwidth}
\includegraphics[width=1.0\textwidth]{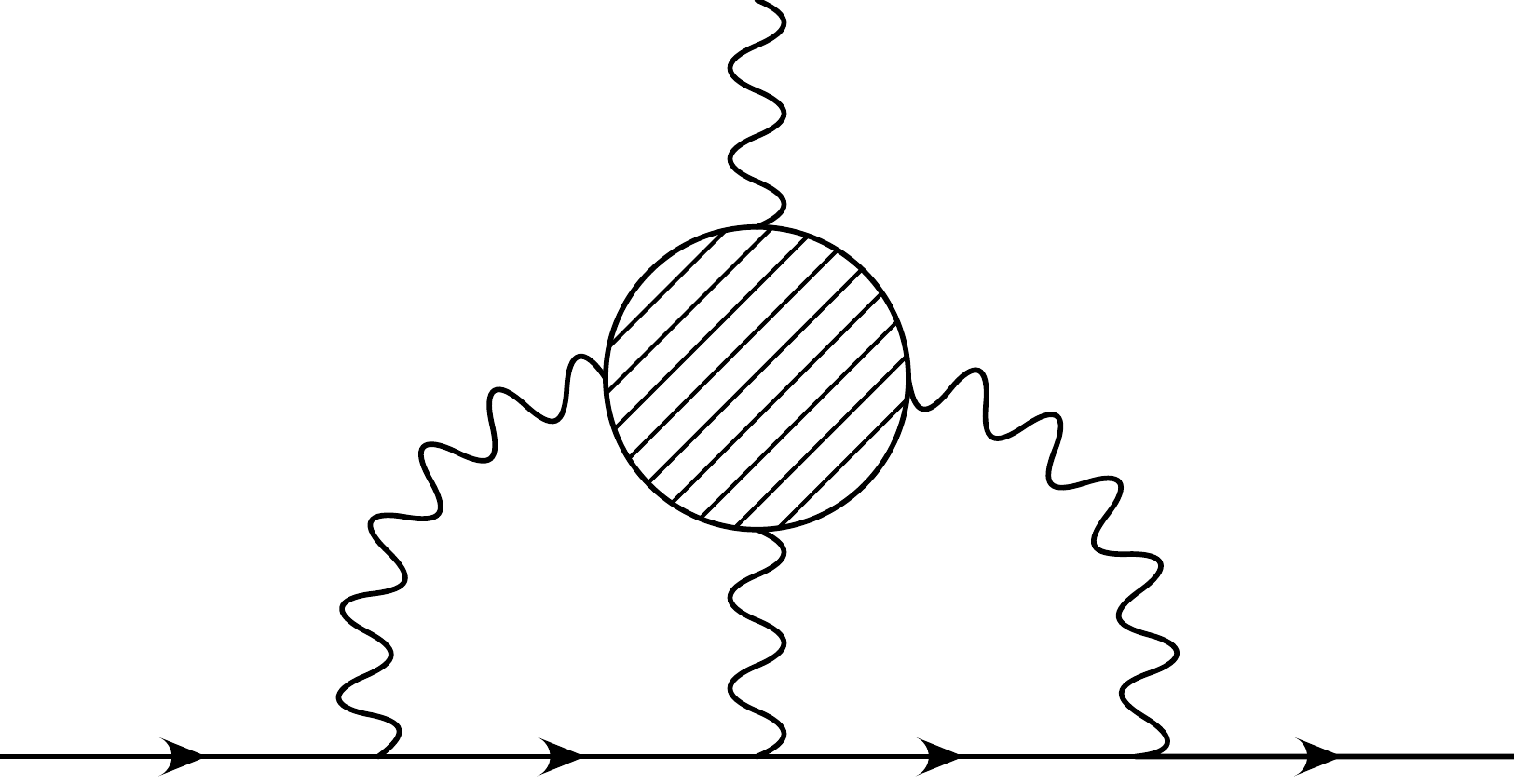} 
\caption{The hadronic light-by-light scattering amplitude. The quark loop represents all possible (``connected'') intermediate hadronic states connected to the horizontal muon line by three photons. The fourth photon represents the external electromagnetic field.}
\label{fig:hlbl}
\end{minipage}\hskip .1\textwidth
\begin{minipage}{0.45\textwidth}
\includegraphics[width=0.9\textwidth]{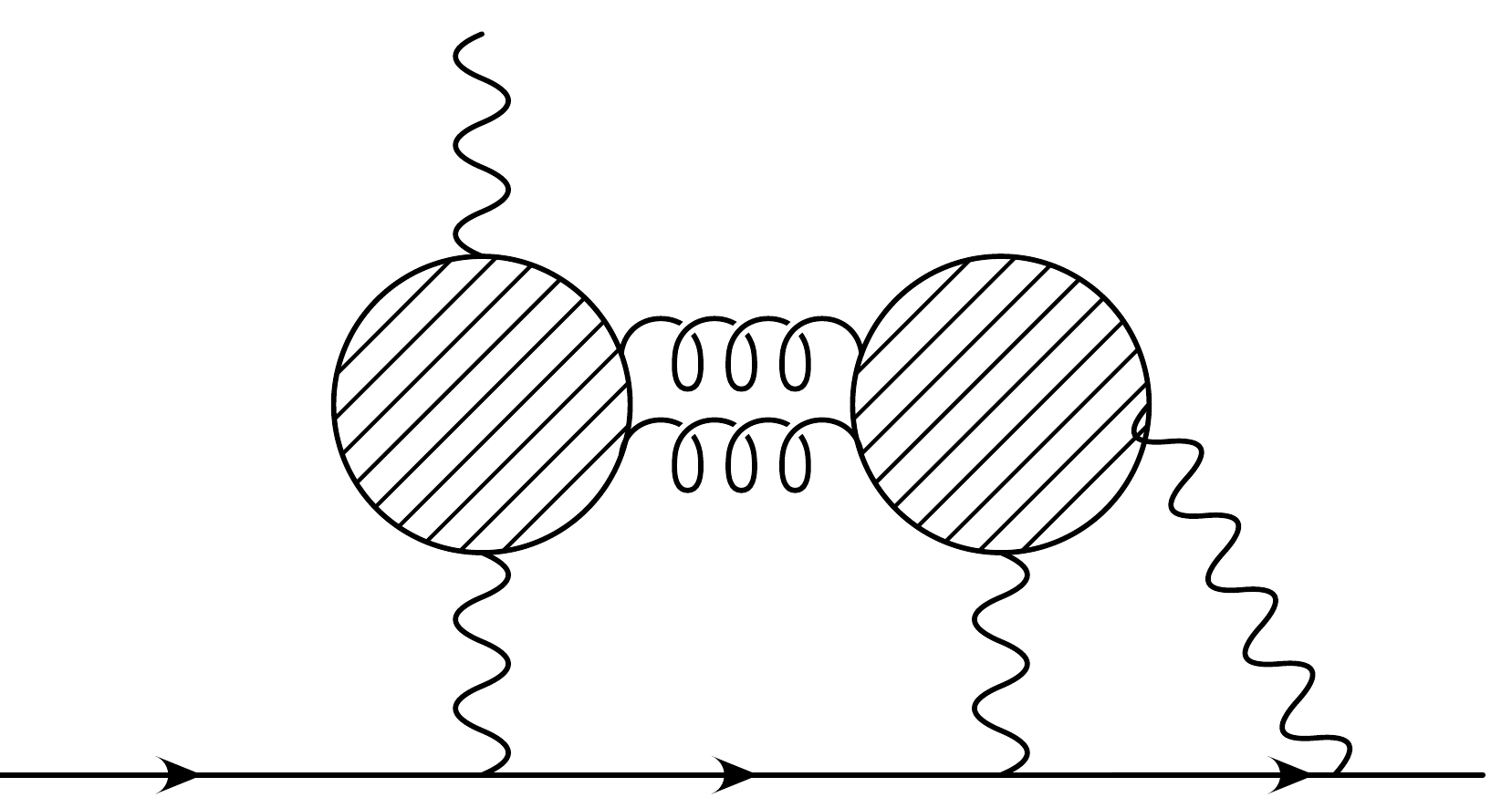} 
\caption{A disconnected quark loop diagram that is part of the hadronic light-by-light scattering amplitude. The quark loops are connected by gluons.}
\label{fig:hlbl disc}
\end{minipage}
\end{figure}

The hadronic light-by-light (HLbL) scattering amplitude is shown in Fig.~\ref{fig:hlbl}. Unlike the HVP contribution, there is no dispersive method of calculation for $a_\mu(\rm HLbL)$. Model estimates put the value at  $a_\mu(\rm HLbL)=10.5(2.6)\times 10^{-10}$~\cite{Prades:2009tw}. The uncertainty quoted for the estimate is part of the "Glasgow consensus" and is not attributable to any one calculation. The errors on the model calculations can not be systematically reduced, and therefore it is desirable to calculate the amplitude directly, using lattice QCD. Notice from Tab.~\ref{tab:summary} that the errors on $a_{\mu}(\rm HVP)$ are roughly two times as large as the quoted model error. Some uncertainties in the literature are larger, for example, see~\cite{Nyffeler:2009tw}. In any case, as the error on the HVP contribution is reduced, the HLbL error will come to dominate the theory error.

The diagram in Fig.~\ref{fig:hlbl} can, in principle, be computed from the correlation of four electromagnetic currents in QCD. Since the aim is to compute the corresponding two-loop QED integral, and one of the electromagnetic currents corresponds to the external photon whose momentum should be taken to zero, two independent momenta are needed for the three remaining vertices due to momentum conservation. This is naively a $(N_t\times N_s^3)^2$ difficult problem. A simpler, but useful exercise, would be to compute the four-point correlation function at some number of fiducial momenta points to check the model calculations. One can also compute a part of the amplitude, like $\pi\to\gamma^{*}\gamma^*$ (where at least on photon is off-shell)~\cite{Feng:2012ck} which is also useful for model calculations.

\begin{figure}[htbp]
\begin{center}
\includegraphics[width=0.4\textwidth]{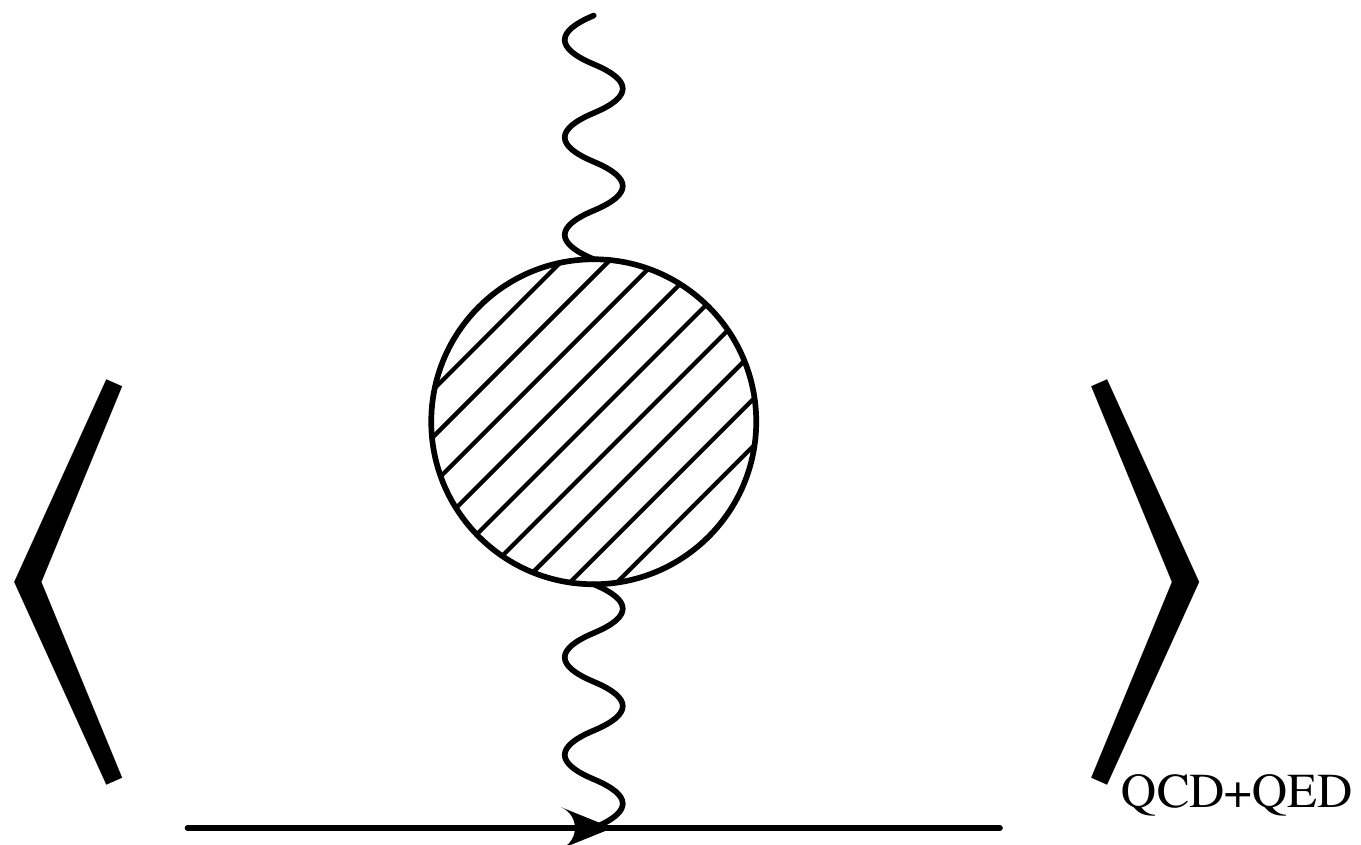} \hskip 0.1\textwidth
\includegraphics[width=0.4\textwidth]{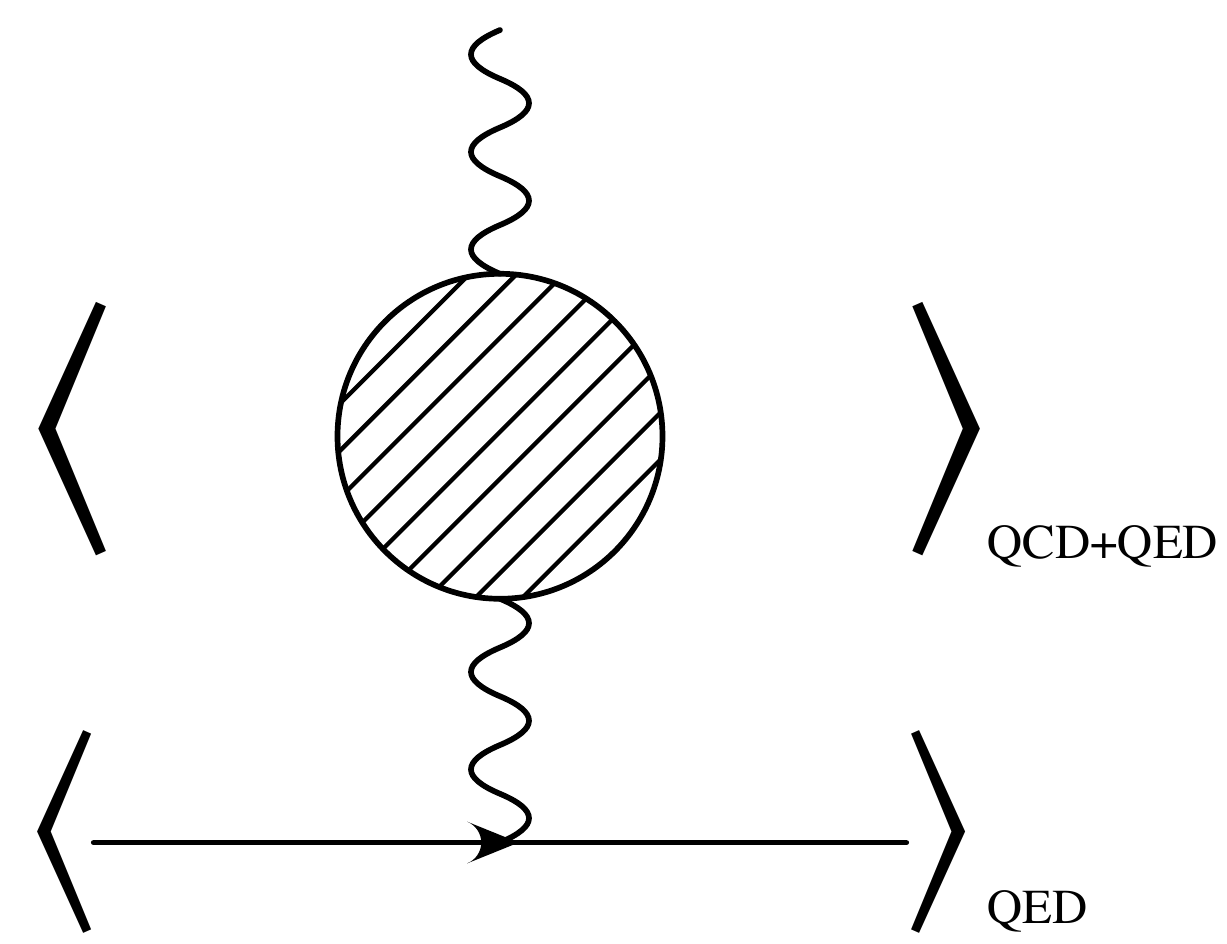} 
\caption{Non-perturbative method for calculating the light-by-light amplitude. The full correlation function with one internal photon inserted by hand is computed in QCD+QED (left). The subtraction term (right) is constructed from separate ensemble averages in QCD+QED and QED. The loop represents the quark (or lepton in pure QED) and the horizontal line is for the muon. The difference of the two diagrams yields the light-by-light scattering amplitude.}
\label{fig:lbl method}
\end{center}
\end{figure}

\begin{figure}[htbp]
\begin{center}
\includegraphics[width=0.9\textwidth]{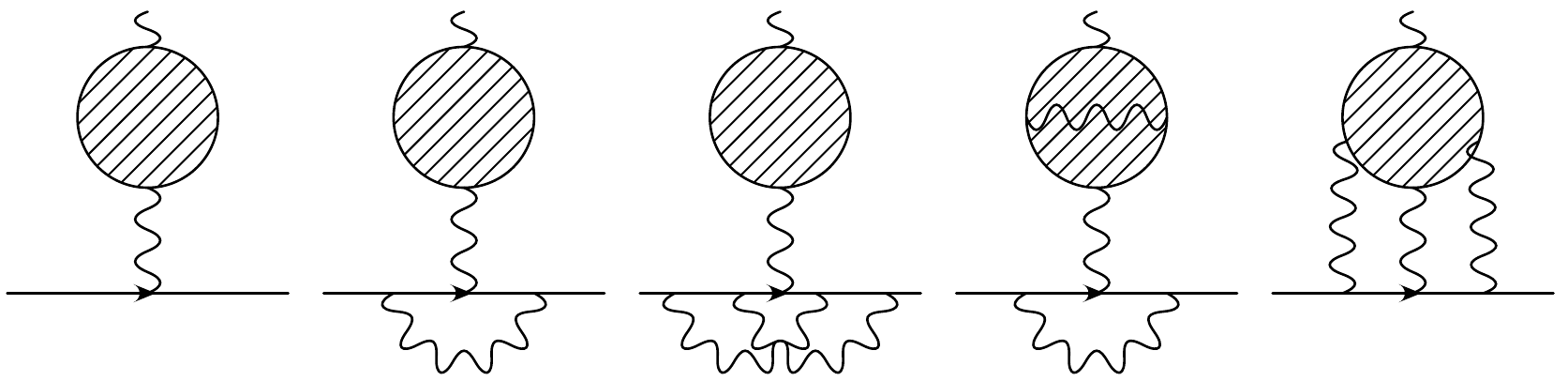}
\caption{Leading and higher order diagrams in the expansion of the correlation and subtraction functions shown if Fig.~\protect\ref{fig:lbl method}. The diagram on the left is $(\alpha)$, the next two are $\alpha^2$, and the last two are $\alpha^3$.}
\label{fig:lbl ho}
\end{center}
\end{figure}

A different approach is to compute the entire amplitude, including the muon and photons, on the lattice~\cite{Hayakawa:2005eq}.   One can either treat the photons in a non-perturbative framework, the procedure that will be followed here, or by treating QED perturbatively but still within the lattice QCD framework. The basic method has been laid out in~\cite{Hayakawa:2005eq} and is shown schematically in Fig.~\ref{fig:lbl method}. 
The idea is to compute correlation functions in a combined photon plus gluon field configuration as first done in~\cite{Duncan:1996xy}. Start by forming a quark loop with two electromagnetic vertices, one for the external photon and another for one of the internal ones. The loop is connected to the muon line by a free lattice photon propagator, so the correlation function is constructed from a two-point quark loop and a three point muon line, where the incoming and outgoing muons have momenta $p$ and $p'$, respectively, and is averaged over the combined QCD+QED gauge field. As it stands, the correlation function has a leading contribution proportional to $\alpha$ (which holds even if the QED configurations are omitted). To extract the HLbL amplitude, which is $O(\alpha^3)$, a non-perturbative subtraction is necessary. Typical higher order terms in the correlation function are shown in Fig.~\ref{fig:lbl ho}. The subtraction term is constructed in exactly the same way, except at the last step, the loop and line are averaged separately over QCD+QED and QED configurations, respectively. The subtraction has exactly the same terms as the original correlation function, through $O(\alpha^{3})$, except the desired HLbL term which is absent since only one photon connects the quark loop and the muon line (see Fig.~\ref{fig:lbl method}). The difference of the two yields the HLbL amplitude. The key point in practice is that both the correlation function and the subtraction term use exactly the same QED and QCD configurations, so they are highly correlated. By using suitable projectors on the subtracted correlation function, linear combinations of the form factors in  Eq.~(\ref{eq:ff}) are obtained ({\it i.e.,} the Sachs form factors $G_{E}$ and $G_{M}$), from which $F_{2}$ is found. 
\begin{eqnarray}
G_E &= & F_1(Q^2) - \frac{Q^2}{(2 m_\mu)^2} F_2(Q^2),\\
G_M &= & F_1(Q^2) + F_2(Q^2).
\end{eqnarray}

In practice, the point-split vector current, which is conserved at non-zero lattice spacing, is used at the internal vertices for the muon line and quark loop so that Ward-Takahashi identities are satisfied exactly, up to numerical precision, on each configuration. These vertices are each Fourier transformed to 4-momentum space, and both the full correlation function and the subtraction correlation function are multiplied by the photon propagator and summed over this 4-momentum. For simplicity, a local current ($\bar\psi\gamma_{\mu}\psi$) which must be renormalized is used for the external vertex.

\subsection{QED contribution}

In order to check the method we started with the calculation in pure QED where the answer is well known in perturbation theory~\cite{Aldins:1969jz,Aldins:1970id}. The calculation was done in quenched non-compact QED, in the Feynman gauge, using domain wall fermions (DWF)~\cite{saumitra}. The lattice size was $16^{3}\times 32$ with $L_{s}=16$ sites in the extra 5th dimension. The muon mass in this test case was relatively large, 0.4 in lattice units, and to enhance the signal the electric charge was set to $e=1.0$ which corresponds to $\alpha=1/4\pi$ instead of 1/137. Here and in the following we always use kinematics where the incoming muon is at rest. 

\begin{figure}[htbp]
\begin{center}
\includegraphics[width=0.4\textwidth]{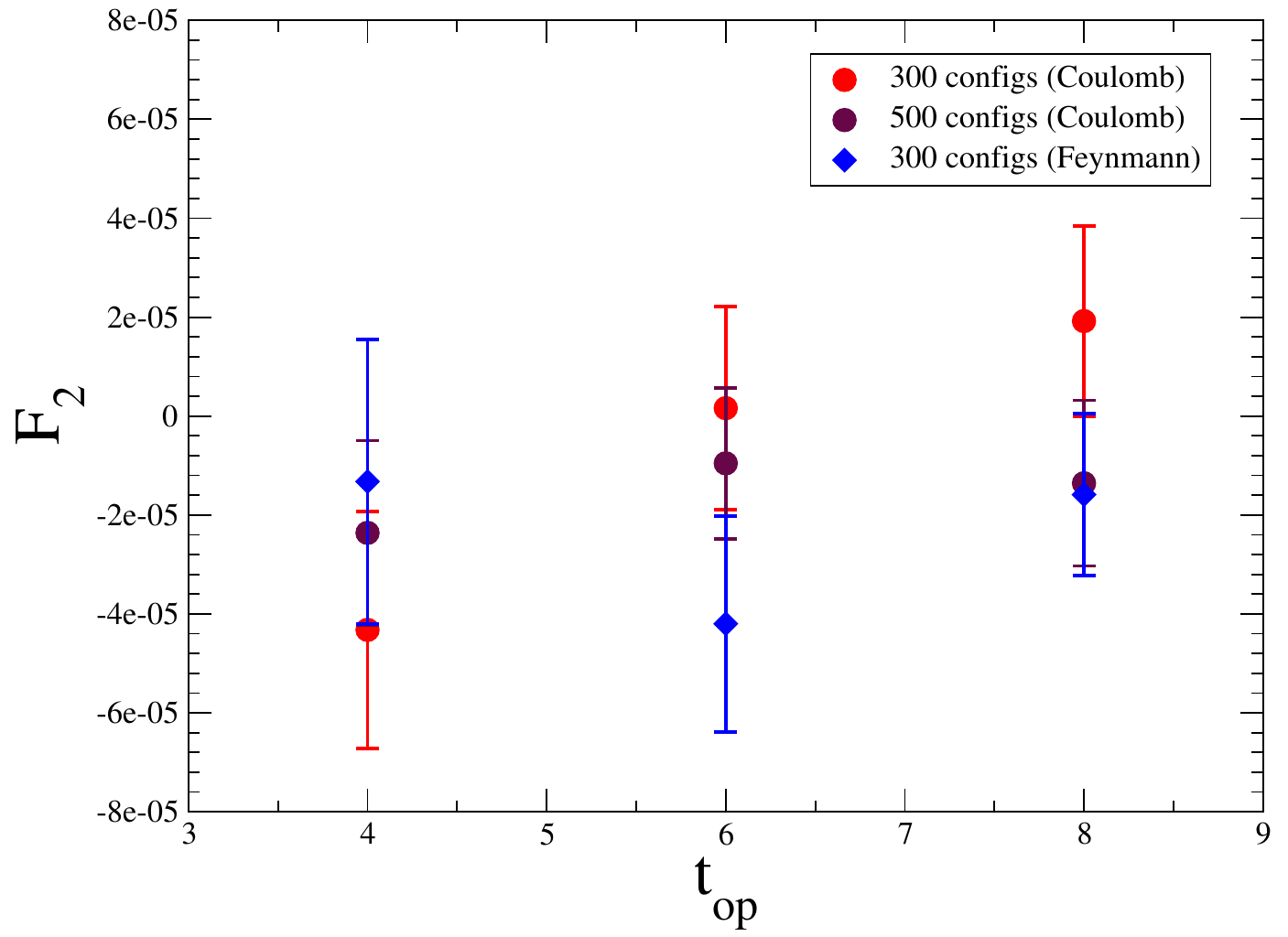} 
\includegraphics[width=0.4\textwidth]{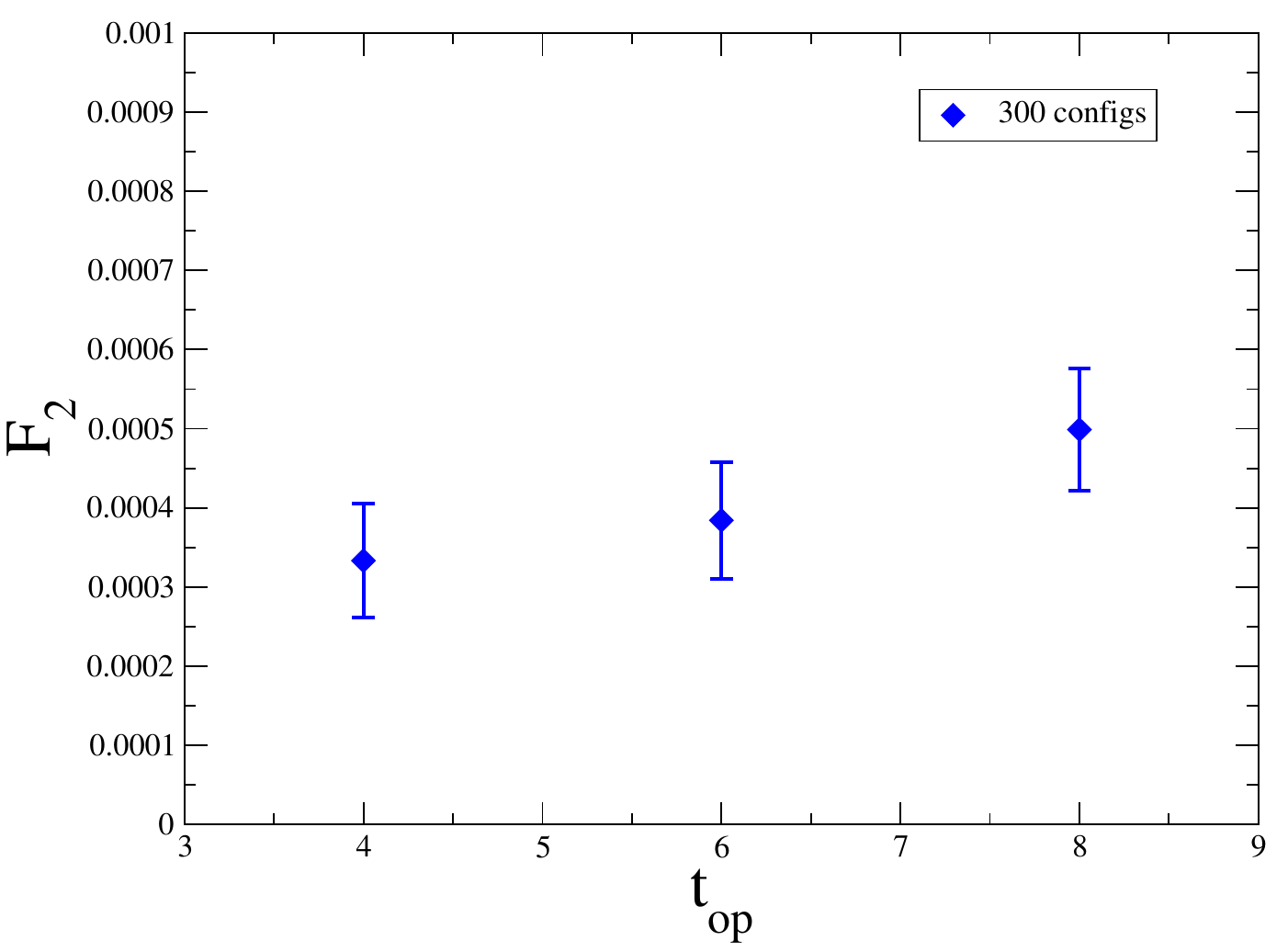} 
\caption{The form factor $F_{2}(Q^2)$ for light-by-light scattering in QED~\cite{saumitra}, evaluated at the lowest non-trivial lattice momenta, $q=2\pi/16$.  The mass of the lepton in the loop is 0.4 (left) and 0.01 (right). $t_{op}$ is the time slice of the external electromagnetic vertex.}
\label{fig:qed f2}
\end{center}
\end{figure}

The first attempt was made with the same heavy mass in the lepton loop as for the muon line, which amounts to computing the electron $g-2$. Using a single point source at the external vertex and relying on plane wave sources for the incoming and outgoing muons to conserve momentum, we were unable to attain a signal (Fig.~\ref{fig:qed f2}), though the statistical error, obtained from a few hundred configurations, was the size of the signal expected from perturbation theory. Taking advantage of the logarithmic enhancement due to the lighter electron mass, $\ln(m_{\mu}/m_{e})$~\cite{Aldins:1970id}, the loop mass was reduced to 0.01. Accordingly, the signal should have increased by an order of magnitude. The result is shown in the right panel of Fig.~\ref{fig:qed f2}; a clear signal is observed. The form factor $F_{2}$ was computed only at the lowest non-trivial momenta, $2 \pi/16$, and was not extrapolated to zero. The average of the three points in the ``plateau'' shown in Fig.~\ref{fig:qed f2} give $F_{2}=3.96(70) \times 10^{-4}=24.4(4.3)(\alpha/\pi)^{3}$ while perturbation theory gives about $10(\alpha/\pi)^{3}$ for $F_2(0)$. The calculation was repeated on a $24^{3}\times32$ lattice to investigate finite volume effects. Again, for the lowest non-trivial momentum, we find $F_{2}=1.19(32)\times 10^{-4}=7.32(1.97)(\alpha/\pi)^{3}$, roughly consistent with perturbation theory. The renormalization factor of the local vector current, inserted at the external vertex, was not included, but its effect which is $O(\alpha)$ should be small compared to other uncertainties.

\subsection{QCD contribution}

The inclusion of QCD into the light-by-light amplitude is straightforward: simply multiply the $U(1)$ gauge links with $SU(3)$ links to create a combined photon and gluon configuration~\cite{Duncan:1996xy}, and follow exactly the same steps, using the same code, as described in the previous sub-section. We use one QED configuration per QCD configuration, though different numbers of each could be beneficial and should be explored. Since the photons are quenched in this first calculation, the valence quarks are charged but not the sea. We discuss the implications of this below.

For QCD we start with the $16^{3}\times 32\times 16$, 2+1 flavor, DWF+Iwasaki ensemble generated by the RBC/UKQCD collaboration~\cite{Blum:2011pu}. The light sea quark mass is 0.01 , the strange 0.032, and the residual mass $m_{\rm res}=0.00308(4)$. The gauge coupling is $\beta=2.13$ and inverse lattice spacing $a^{-1}=1.73(3)$ GeV. The corresponding pion mass is $m_\pi=422$ MeV, and the muon mass is $m_{\mu}=0.4$ \footnote{Using the QCD lattice spacing, this corresponds to 692 MeV compared to the physical mass, $m_{\mu}=105.658367(4)$.} is the same as in the pure QED case. The external electromagnetic vertex is inserted on time slice $t_{\rm op}=6$ and the incoming and outing muons are created and destroyed at $t=0$ and 12, respectively. The incoming muon is again at rest while the outgoing muon has one unit of momentum (we average results over all three directions, and $\pm \vec p$). Finally, $e=1$ as before. While these unphysical parameters may induce relatively large systematic errors, the main aim of this study is to show that the statistical errors can  be controlled and that the method works. After demonstrating this, the systematic errors can be addressed.

In the QED case, setting $m_{e}\ll m_{\mu}$ lead to a large enhancement of the signal to noise ratio. The same effect is not expected here since $m_{\pi}\sim m_{\mu}$. Even in the physical case, $m_{\pi}/m_{\mu}\approx 1.33$ which may partly explain why the HLbL amplitude is about 200 times smaller than the QED amplitude. To reduce the statistical error we use a point source at the external vertex of the quark loop every 4 spatial sites in each direction, for a total of 64 propagator calculations per configuration (100 combined-configurations were used). 

Following the same procedure as in the pure QED case, we find $F_{2}(Q^2=0.38\,\rm GeV^{2})= (-15.7\pm2.3) \times 10^{-5}$. While the signal is robust, the magnitude is between 5-10 times larger than the model estimates for $a_{\mu}(\rm HLbL)$, and even has the opposite sign. There are several reasons why this might be the case. First, models are not expected to be valid in the range of masses used here. Since the pion mass is heavy, the leading pion pole term is probably not dominant and the sub-leading terms can be important~\cite{Prades:2009tw}. Some of them have opposite signs~\cite{Prades:2009tw}. Further, the sub-leading terms are known to be proportional to $m_{\mu}^{2}$~\cite{Prades:2009tw}, enhancing the effect of the large muon mass used here.

An important check that the subtraction is working is to vary $e$ and see that the amplitude changes in the expected proportion compared to the case where $e=1$. The same non-compact QED configurations are used in each case; $e$ is varied only when constructing the exponentiated gauge-link, $U_{\mu}(x) = \exp{ie A_{\mu}(x)}$. Thus the ratio of form factors, and hence $\alpha$ dependence, can be determined very accurately. Since one photon is inserted explicitly, and the charge at this vertex is not included in the lattice calculation, the subtracted amplitude should behave like $\sim e^{4}$. Using $e=0.84$ and 1.19, the changes relative to $e=1$ should be 0.5 and 2.0, respectively. This is precisely what is observed. In addition, there is no detectable change in the unsubtracted amplitudes.

The next step, which is not too costly, is to reduce only the muon mass. With $m_{\mu}=0.1$ and everything else unchanged, except that the number of configurations was increased to 306,  we find $F_{2}(Q^2=0.19\,\rm GeV^{2})= -2.21\pm0.75 \times 10^{-5}$. This reduction is consistent with the already mentioned leading $m_{\mu}^{2}$ dependence. Note the value of $Q^{2}$ is also a factor of two smaller because of the smaller muon mass.

Our next step was to increase the lattice size from $16^{3}\times 32$ to $24^{3}\times 64$ and reduce the quark mass to 0.005 (also an RBC/UKQCD 2+1 flavor, DWF+Iwasaki ensemble~\cite{Allton:2008pn,Aoki:2010dy}). The pion mass measured on this ensemble is 329 MeV, and volume is increased by more than a factor of three.  The increased lattice size and smaller quark mass lead us to use the all mode averaging (AMA) technique~\cite{Blum:2012uh} to maintain large statistics at an affordable cost. Besides the exact calculation which was done using a single point source on 20 configurations, the approximation was computed using 400 lowmodes of the even-odd preconditioned Dirac operator and 216 point sources computed with stopping residual $10^{-4}$ on 157 configurations\footnote{This is about double the statistics as were presented at the meeting.}. The external vertex was inserted on time slice $t_{\rm op}=5$ with incoming and outgoing muons at 0 and 9, respectively. The incoming muon is at rest, while the outgoing has three-momenta in units of $2\pi/L$ of $(\pm1,0,0)$ and $(\pm 1,\pm1,0)$ and permutations. We also include the vector current renormalization (in pure QCD) from~\cite{Aoki:2010dy}.

\begin{figure}[htbp]
\begin{center}
\includegraphics[width=0.8\textwidth]{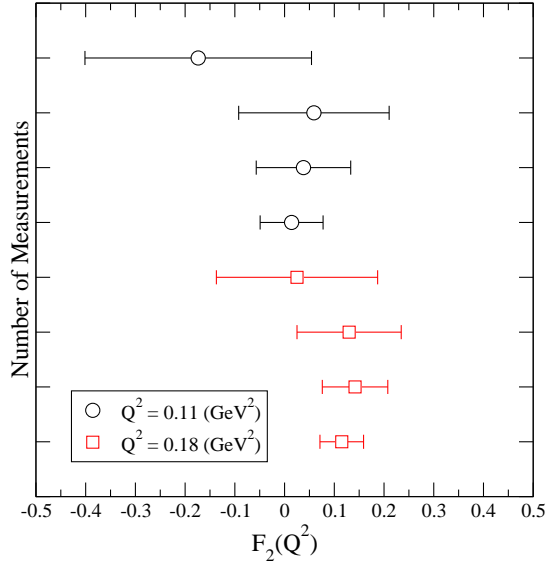} 
\caption{$F_{2}(Q^2)$ computed on the $24^{3}$, $m_{\pi}=329$ MeV ensemble. Lowest two momenta only. The statistical errors scale with $1/\sqrt{N}$ where $N$ is the number of measurements which range from 20 to 157, roughly doubled at each step. The central value is also stable, and the size of model estimates.}
\label{fig:stability}
\end{center}
\end{figure}

In Fig.~\ref{fig:stability} we show the stability of the signal for $F_{2}(Q^2)$ for the two lowest non-trivial momenta ($Q^2=0.11$ and 0.18 GeV$^{2}$) as more configurations are added to the ensemble average (20 to 157, roughly doubling at each step). The statistical errors scale like $1/\sqrt{N}$, where $N$ is the number of measurements. The central value is stable, and has magnitude comparable in the size to model estimates (and statistical errors only 2-3 times larger). The sign is also the same.
\begin{eqnarray}
F_2(0.18{\rm ~GeV}^2)&=&(0.115 \pm 0.044)\times \left(\frac{\alpha}{\pi}\right)^3,\label{eq:f2 q1}\\
F_2(0.11{\rm ~GeV}^2)&=&(0.014 \pm 0.063)\times \left(\frac{\alpha}{\pi}\right)^3,\label{eq:f2 q2}\\
a_\mu({\rm HLbL/model})&=&(0.084 \pm 0.020)\times \left(\frac{\alpha}{\pi}\right)^3~\cite{Prades:2009tw}.
\end{eqnarray}

Several comments are in order. The signal for the larger $Q^2$ point is about 3 standard deviations while the lowest $Q^2$ point is zero within the statistical error. The former has more momenta combinations to average over, 12 versus 6, which is reflected in the errors. The fact that the sign is consistent with model estimates (and is the same as for pure QED) may be due to the reduced pion mass (pion pole term becoming dominant), or the larger volume, or a combination of the two. In the AMA procedure~\cite{Blum:2012uh} the expectation value of an operator is given by
\begin{eqnarray}
\langle {\cal O} \rangle &=& \langle {\cal O}_{\rm rest}\rangle+\frac{1}{N_G}\sum_{g}\langle {\cal O}_{\rm approx,g}\rangle,
\label{eq:ama}
\end{eqnarray}
where $N_G$ is the number of measurements of the approximate observable, and ``rest" refers to the contribution of the exact observable minus the approximation, evaluated for the same conditions. In the present calculation, the central value and errors appearing in Eqs.~(\ref{eq:f2 q1}) and (\ref{eq:f2 q2}) are completely dominated by the second term on the r.h.s. of Eq.~(\ref{eq:ama}).

We briefly comment on the absence of sea-quark electric charge in our calculation. The QED quenching of the sea-quarks is equivalent to omitting diagrams like the one shown in Fig.~\ref{fig:hlbl disc}, and variants with three quark loops, or ones generated by interchanging loops containing the external vertex. In principle these could be calculated in the quenched setup, by introducing one and two additional valence quark loops. However it is straightforward to include them by repeating the calculation above, but on dynamical QCD+QED configurations, or equivalently, by re-weighting the quenched QED configurations~\cite{Ishikawa:2012ix}. We also note that there is no reason, {\it a-priori}, to expect that the size of the omitted diagrams is any smaller than what has already been calculated, except possibly by Zweig suppression. 

\section{Conclusions}
\label{sec:con}

The muon anomalous magnetic moment continues to provide one of the most precise tests of the Standard Model of particle physics and often places important constraints on new theories beyond the Standard Model. With new experiments planned at Fermilab (E989) and J-PARC (E34) that aim to improve on the current 0.54 ppm measurement at BNL by at least a factor of four, it will continue to play a central role in particle physics for the foreseeable future. 

Owing to the non-perturbative nature of QCD, the hadronic corrections to the muon $g-2$ are the largest source of error in the Standard Model calculation. These errors must be reduced to leverage the new experiments~\cite{Hewett:2012ns}. The hadronic corrections enter at order $\alpha^{2}$ through the hadronic vacuum polarization (HVP) and $\alpha^{3}$ through hadronic light-by-light (HLbL) scattering as well as higher order HVP contributions. Both are being calculated using lattice QCD.

While the HVP contribution to the muon anomaly has been precisely computed to the 0.6\% level using experimental measurements of $e^{+}e^{-}\to\rm hadrons$ and $\tau\to\rm hadrons$, lattice QCD calculations serve as an important independent check of these results. At the moment, statistical errors on lattice calculations of $a_{\mu}(\rm HVP)$ are at about the 3-5\% level, but important systematic errors remain. Most significant is that for light quark masses the errors on the low-momentum region of $\Pi(Q^2)$ are not small enough, nor are there sufficient points available in the crucial region, $Q^2\sim m_{\mu}^{2}$, to adequately estimate $a_{\mu}(\rm HVP)$. Quark masses are too heavy (or errors are still too large for light masses) and fits are model-dependent.
The good news is that all of these points are being addressed in the newest calculations. Lattice calculations using model independent fit functions, noise reduction techniques, twisted boundary conditions, charm quarks, and physical light quark masses on large lattices are underway: Big error reductions over the next one to two years are not only possible, but likely. Finally, to get to the 1\% level, or better, disconnected diagrams and isospin breaking effects must be incorporated to complete the calculation.

The HLbL contribution to the muon anomaly can not be computed using data from experiment and a dispersion relation as for the HVP contributions.
It is expected to dominate the theory error as the HVP error is reduced (either from more experimental data, lattice calculations, or both). 
We have presented encouraging first results for the part of the HLbL contribution where there is a single quark loop in the corresponding diagram. Calculations in pure QED were used to successfully check the method.  The calculation is numerically intensive, but logically straightforward. Much effort is still needed to reduce statistical errors, extrapolate to zero momentum transfer, and many systematic errors remain uncontrolled. The calculation is quenched with respect to QED, so the sea quarks are not charged, and potentially large contributions are missing. This can be fixed by using dynamical QCD+QED gauge configurations, or by re-weighting the quenched QED ensembles.

\section*{Acknowledgements} We thank USQCD and the RIKEN BNL Research Center for computing resources used in the HLbL work. M.H. is supported in part by Grants-in-Aid for Scientific Research (S)22224003, (C)20540261, TB is supported in part by the US Department of Energy under Grant No. DE-FG02-92ER40716, and TI is supported in part by Grants-in-Aid for Scientific Research 22540301 and 23105715 and under U.S. DOE grant DE-AC02-98CH10886.


\begin{thebibliography}{99}
\bibitem{Bennett:2006fi} 
  G.~W.~Bennett {\it et al.}  [Muon G-2 Collaboration],
  Phys.\ Rev.\ D {\bf 73}, 072003 (2006)
  [hep-ex/0602035].
\bibitem{Aoyama:2012wk} 
  T.~Aoyama, M.~Hayakawa, T.~Kinoshita and M.~Nio,
  Phys.\ Rev.\ Lett.\  {\bf 109}, 111808 (2012)
  [arXiv:1205.5370 [hep-ph]].
\bibitem{Davier:2010nc} 
  M.~Davier, A.~Hoecker, B.~Malaescu and Z.~Zhang,
  Eur.\ Phys.\ J.\ C {\bf 71}, 1515 (2011)
  [Erratum-ibid.\ C {\bf 72}, 1874 (2012)]
  [arXiv:1010.4180 [hep-ph]].
\bibitem{Hagiwara:2011af} 
  K.~Hagiwara, R.~Liao, A.~D.~Martin, D.~Nomura and T.~Teubner,
  J.\ Phys.\ G {\bf 38}, 085003 (2011)
  [arXiv:1105.3149 [hep-ph]].
\bibitem{Czarnecki:2002nt} 
  A.~Czarnecki, W.~J.~Marciano and A.~Vainshtein,
  Phys.\ Rev.\ D {\bf 67}, 073006 (2003)
  [Erratum-ibid.\ D {\bf 73}, 119901 (2006)]
  [hep-ph/0212229].
\bibitem{Hagiwara:2006jt} 
  K.~Hagiwara, A.~D.~Martin, D.~Nomura and T.~Teubner,
  Phys.\ Lett.\ B {\bf 649}, 173 (2007)
  [hep-ph/0611102].
\bibitem{Hewett:2012ns} 
  J.~L.~Hewett, H.~Weerts, R.~Brock, J.~N.~Butler, B.~C.~K.~Casey, J.~Collar, A.~de Govea and R.~Essig {\it et al.},
  arXiv:1205.2671 [hep-ex].
\bibitem{Blum:2002ii} 
  T.~Blum,
  Phys.\ Rev.\ Lett.\  {\bf 91}, 052001 (2003)
  [hep-lat/0212018].
\bibitem{Prades:2009tw} 
  J.~Prades, E.~de Rafael and A.~Vainshtein,
  (Advanced series on directions in high energy physics. 20)
  [arXiv:0901.0306 [hep-ph]].
\bibitem{Durand:1962zzb} 
  L.~Durand,
  Phys.\ Rev.\  {\bf 128}, 441 (1962).
\bibitem{Gourdin:1969dm} 
  M.~Gourdin and E.~De Rafael,
  Nucl.\ Phys.\ B {\bf 10}, 667 (1969).
\bibitem{Alemany:1997tn} 
  R.~Alemany, M.~Davier and A.~Hocker,
  Eur.\ Phys.\ J.\ C {\bf 2}, 123 (1998)
  [hep-ph/9703220].
\bibitem{Jegerlehner:2011ti} 
  F.~Jegerlehner and R.~Szafron,
  Eur.\ Phys.\ J.\ C {\bf 71}, 1632 (2011)
  [arXiv:1101.2872 [hep-ph]].
\bibitem{Lautrup:1971yp} 
  B.~E.~Lautrup, A.~Peterman and E.~De Rafael,
  Nuovo Cim.\ A {\bf 1}, 238 (1971).
\bibitem{Chetyrkin:1996cf} 
  K.~G.~Chetyrkin, J.~H.~Kuhn and M.~Steinhauser,
  Nucl.\ Phys.\ B {\bf 482}, 213 (1996)
  [hep-ph/9606230].
\bibitem{Renner:2012fa} 
  D.~B.~Renner, X.~Feng, K.~Jansen and M.~Petschlies,
  arXiv:1206.3113 [hep-lat].
\bibitem{Gockeler:2003cw} 
  M.~Gockeler {\it et al.}  [QCDSF Collaboration],
  Nucl.\ Phys.\ B {\bf 688}, 135 (2004)
  [hep-lat/0312032].
\bibitem{Aubin:2006xv} 
  C.~Aubin and T.~Blum,
  Phys.\ Rev.\ D {\bf 75}, 114502 (2007)
  [hep-lat/0608011].
  \bibitem{Aubin:2012}
  C. Aubin, \href{https://indico.fnal.gov/contributionDisplay.py?sessionId=9&contribId=56&confId=5276}{talk} at the Project X Physics Summer Study workshop, Fermilab, June, 2012.
\bibitem{Boyle:2011hu} 
  P.~Boyle, L.~Del Debbio, E.~Kerrane and J.~Zanotti,
  Phys.\ Rev.\ D {\bf 85}, 074504 (2012)
  [arXiv:1107.1497 [hep-lat]].
\bibitem{Feng:2011zk} 
  X.~Feng, K.~Jansen, M.~Petschlies and D.~B.~Renner,
  Phys.\ Rev.\ Lett.\  {\bf 107}, 081802 (2011)
  [arXiv:1103.4818 [hep-lat]].
\bibitem{DellaMorte:2011aa} 
  M.~Della Morte, B.~Jager, A.~Juttner and H.~Wittig,
  JHEP {\bf 1203}, 055 (2012)
  [arXiv:1112.2894 [hep-lat]].
\bibitem{DellaMorte:2012cf} 
  M.~Della Morte, B.~Jager, A.~Juttner and H.~Wittig,
  PoS LATTICE {\bf 2012}, 175 (2012)
  [arXiv:1211.1159 [hep-lat]].
\bibitem{Feng:2012gh} 
  X.~Feng, G.~Hotzel, K.~Jansen, M.~Petschlies and D.~B.~Renner,
  arXiv:1211.0828 [hep-lat].
\bibitem{Aubin:2012me} 
  C.~Aubin, T.~Blum, M.~Golterman and S.~Peris,
  Phys.\ Rev.\ D {\bf 86}, 054509 (2012)
  [arXiv:1205.3695 [hep-lat]].
\bibitem{Golterman} M. Golterman, these proceedings.
\bibitem{Shintani} E. Shintani, these proceedings.
\bibitem{Blum:2012uh}
  T.~Blum, T.~Izubuchi and E.~Shintani,
  arXiv:1208.4349 [hep-lat].
\bibitem{DellaMorte:2010aq} 
  M.~Della Morte and A.~Juttner,
  JHEP {\bf 1011}, 154 (2010)
  [arXiv:1009.3783 [hep-lat]].
\bibitem{Aubin} C. Aubin and T. Blum, unpublished data for the HVP
calculated on MILC ensemble l48144f21b747m0036m018.
\bibitem{Nyffeler:2009tw} 
  A.~Nyffeler,
  Phys.\ Rev.\ D {\bf 79}, 073012 (2009)
  [arXiv:0901.1172 [hep-ph]].
\bibitem{Feng:2012ck} 
  X.~Feng, S.~Aoki, H.~Fukaya, S.~Hashimoto, T.~Kaneko, J.~-i.~Noaki and E.~Shintani,
  arXiv:1206.1375 [hep-lat].
\bibitem{Hayakawa:2005eq} 
  M.~Hayakawa, T.~Blum, T.~Izubuchi and N.~Yamada,
  PoS LAT {\bf 2005}, 353 (2006)
  [hep-lat/0509016].
\bibitem{Duncan:1996xy}
  A.~Duncan, E.~Eichten and H.~Thacker,
  Phys.\ Rev.\ Lett.\  {\bf 76} (1996) 3894
  [hep-lat/9602005].
\bibitem{Aldins:1969jz} 
  J.~Aldins, T.~Kinoshita, S.~J.~Brodsky and A.~J.~Dufner,
  Phys.\ Rev.\ Lett.\  {\bf 23}, 441 (1969).
\bibitem{Aldins:1970id} 
  J.~Aldins, T.~Kinoshita, S.~J.~Brodsky and A.~J.~Dufner,
  Phys.\ Rev.\ D {\bf 1}, 2378 (1970).
\bibitem{saumitra}
S. Chowdhury, Ph.D. thesis, University of Connecticut, 2009.
\bibitem{Blum:2011pu} 
  T.~Blum, P.~A.~Boyle, N.~H.~Christ, N.~Garron, E.~Goode, T.~Izubuchi, C.~Lehner and Q.~Liu {\it et al.},
  Phys.\ Rev.\ D {\bf 84}, 114503 (2011)
  [arXiv:1106.2714 [hep-lat]].
\bibitem{Allton:2008pn} 
  C.~Allton {\it et al.}  [RBC-UKQCD Collaboration],
  Phys.\ Rev.\ D {\bf 78}, 114509 (2008)
  [arXiv:0804.0473 [hep-lat]].
\bibitem{Aoki:2010dy} 
  Y.~Aoki {\it et al.}  [RBC and UKQCD Collaborations],
  Phys.\ Rev.\ D {\bf 83}, 074508 (2011)
  [arXiv:1011.0892 [hep-lat]].
\bibitem{Ishikawa:2012ix} 
  T.~Ishikawa, T.~Blum, M.~Hayakawa, T.~Izubuchi, C.~Jung and R.~Zhou,
  Phys.\ Rev.\ Lett.\  {\bf 109}, 072002 (2012)
  [arXiv:1202.6018 [hep-lat]].
\end{thebibliography}
\end{document}